\documentclass[%
 reprint,
 amsmath,amssymb,
 aps,
pra,
]{revtex4-2}

\usepackage{graphicx}
\usepackage{dcolumn}
\usepackage{bm}
\usepackage{comment}
\usepackage{color}
\usepackage{soul}

\usepackage{graphicx}
\usepackage{dcolumn}
\usepackage{bm}
\usepackage{comment}
\usepackage{color}
\usepackage{soul}
\usepackage{amsmath,amssymb}

\begin{document}








\title{Localization Tensor Revisited: Geometric–Probabilistic Foundations and a Structure-Factor Criterion under Periodic Boundaries}

\author{Zhe-Hao Zhang$^{1}$}
\author{Xiaoming Cai$^{2}$}
\author{Yi-Cong Yu$^{2}$}\thanks{ycyu@wipm.ac.cn}

\affiliation{$^{1}$Institute for Advanced Study in Physics and School of Physics, Zhejiang University, Hangzhou 310058, China}
\affiliation{$^{2}$
Innovation Academy for Precision Measurement Science and Technology, 
Chinese Academy of Sciences, Wuhan 430071, China}

\begin{abstract}
We revisit the localization tensor (LT) from geometric and probabilistic perspectives and construct extensions that are naturally compatible with periodic boundary conditions (PBC), without redefining the position operator. In open boundary conditions, we show that the LT can be written exactly as the covariance of a bivariate probability distribution built from density--density correlations. This leads to two conceptually distinct extensions to PBC: (i) a geometric one based on the Riemannian center (Fr\"echet mean) on the circle, and (ii) a metric-free one based on the mutual information
\(I\), which treats the configuration space purely as a probability space.
We then relate the LT to the static structure factor by identifying the diagonal part, \(C_{pp}\), as a ``localization function'' \(C(p)\), whose small-momentum behavior determines the LT in the thermodynamic limit. This clarifies why the LT is sensitive to transitions out of the extended phase but by itself cannot distinguish Anderson-type localization from dimerization: both share the same low-momentum asymptotics. We show that the finite-momentum behavior of \(C(p)\), together with an inverse participation ratio (IPR) based upper bound valid in localized phases, provides a sharp criterion that discriminates localization from dimerization. These results are illustrated on the Su--Schrieffer--Heeger and Aubry--Andr\"e models, with and without interactions, and suggest that structure factor based probes offer robust and experimentally accessible diagnostics of localized and dimerized phases under PBC.
\end{abstract}

\maketitle

\section{\label{sec:introduction}Introduction}
Localization is an important concept in condensed matter physics, closely related to the conductive and transport properties of materials \cite{mott1984,roati2008,mendoza2016}. In localization studies, the position of electrons (or other microscopic particles) is a significant research focus. Traditionally, the description of the positional information of a quantum mechanical system relies on the calculation involving the position operator \cite{griffiths2018,landau2013}. It is regarded that  the spread of the position operator is able to discriminate between systems that behave as insulators or conductors in the thermodynamic limit\cite{kohn1964,souza2000,resta1998,resta1999,resta2007}. However, in certain situations, performing calculations directly using the position operator may lead to some problem, especially under periodic boundary conditions (PBC) \cite{valencaferreiradearagao2019,evangelisti2022,ditler2023}. It is not a peculiar phenomenon, as it is evident that $x\psi(x)$ is no longer a periodic function while $\psi(x)$ itself is periodic, this implies that the usual position operator (in the function representation) defined by $\hat{X} := x $ does not remain closed within the Hilbert space constituted by periodic functions. This problem has been recognized for a long time, and several solutions have been proposed \cite{otto1992,kudinov1999,ditler2023}.  In practical calculations, a simple redefinition of the position operator is computationally friendly and widely adopted \cite{azor2021,alves2021,escobarazor2019, escobarazor2024, tavernier2020, tavernier2021, brumas2024, alrakik2025}.

Research on the redefinition of the position operator in recent years has its roots in the conclusions drawn from the study of polarization problems in condensed matter physics in the 1990s \cite{otto1992,ortiz1994,king-smith1993}. Resta and other researches, based on the microscopic theory of quantum mechanics, discovered through calculations that the contribution to the polarization of electrons is a topologically related quantity \cite{zak1989, thouless1983}
$ \boldsymbol{P}_\mathrm{ele} \propto -\mathrm{i}\sum_n\int \mathrm{d} \boldsymbol{k}\boldsymbol{A}(\boldsymbol{k})$ ,here $\boldsymbol{A}(\boldsymbol{k})$ is the Berry connection defined by $\boldsymbol{A}(\boldsymbol{k})=\mathrm{i}\langle  u_{n\boldsymbol{k}}\vert \nabla_{\boldsymbol{k}}  \vert u_{n\boldsymbol{k}} \rangle$ and $\vert u_{n\boldsymbol{k}} \rangle$ denotes the Bloch wave function with $n$ being the band index and $\boldsymbol{k}$ the Bloch wave vector \cite{resta1994, resta1998}. Based on this observation, Resta arrive at the central result of the modern theory of polarization \cite{resta2007}: the formal polarization is only well-defined $\mathit{mod}$ $e\boldsymbol{R}/\Omega$, where $e$ is the electron charge, $\boldsymbol{R}$ is any lattice vector and $\Omega$ is the primitive-cell volume. In subsequent research, this problem was more clearly elucidated. Resta, after introducing the periodicity $L$, derived the expression for the electronic polarizability in terms of the position operator $\hat{X} = \sum x_i$ and the ground-state wave function $\vert \psi_0 \rangle$  in the thermodynamic limit \cite{resta1998}
\begin{equation}
\boldsymbol{P}_\mathrm{ele} = \lim_{L \to \infty} \frac{e}{2\pi}\mathrm{Im} \left[\ln\langle \psi_0 \vert \exp (\mathrm{i}\frac{2\pi}{L}\hat{X} )\vert \psi_0 \rangle \right].
\label{eqPele}
\end{equation}
where $\mathrm{i}$ the imaginary unit, $L$ is the period and $\vert \psi_0 \rangle$ is the ground state electronic wave function. By comparing the result of equation (\ref{eqPele}) with the definition of electronic polarizability in the classical picture given by $\boldsymbol{P} = \Omega^{-1} \int_\Omega \mathrm{d} \boldsymbol{r} \, \boldsymbol{r} \rho (\boldsymbol{r})$, he arrived at a reasonable hypothesis for the expectation value of the position operator
\cite{resta1998}:
\begin{equation}
\langle x \rangle := \frac{L}{2\pi}\mathrm{Im} \left[\ln\langle \psi_0 \vert \exp (\mathrm{i}\frac{2\pi}{L}\hat{X} )\vert \psi_0 \rangle \right].
\label{eqRestaDef}
\end{equation}

From the current perspective, equation (\ref{eqRestaDef}) can be regarded as a refined expression of a Wilson loop \cite{alexandradinata2014, wang2019}.  To clarify this point (we will now discuss the problem in one dimension, this discussion can be readily generalized to arbitrary dimensions), let us first assume that the ground state $\vert \psi_0 \rangle$ is constructed by filling the following orbitals \cite{hetenyi2019}:
\begin{align}
    \vert \psi_0 \rangle = \mathcal{A} \prod_{(m,k) \in \text{occ}} \phi_{m,k},
    \label{eqPsi0}
\end{align}
here, $m$ denotes the band index and $k$ denotes the Bloch wave vector, $\mathcal{A}$ is the total anti symmetrization operator, which makes $\psi_0$ a slater determinant. The action of the operator $\exp(\mathrm{i}\frac{2\pi}{L}\hat{X})=\exp (\mathrm{i}\frac{2\pi}{L}\sum_n x_n)$ on the state $\vert \psi_0 \rangle$ is equivilent to performing a transformation on each orbit $\tilde{\phi}_{m,k} = \exp(\mathrm{i}\frac{2\pi}{L}x)\phi_{m,k}$. Therefore, the part inside the logarithm on the right-hand side of equation (\ref{eqRestaDef}) can be expressed as a determinant:
\begin{align}
    &\langle \psi_0 \vert \exp(\mathrm{i}\frac{2\pi}{L}\hat{X})\vert \psi_0 \rangle = \det S,
    \notag \\
    & S_{m,k ; m^\prime, k^\prime} :=
    \langle \phi_{m, k} \vert \exp(\mathrm{i}\frac{2\pi}{L}x) \vert \phi_{m^\prime,k^\prime} \rangle.
    \label{eqDetS}
\end{align}
Note that since $k$-space has already been discretized by $\Delta k=2\pi/L$, the matrix $S$ is block sub-diagonal, its elements are nonzero only when $k = k^\prime + 2 \pi/L$. This allow us to express $\det S$ as a product form:
\begin{align}
    \det S = \prod_k \det S(k, k-\Delta k), 
    \label{eqDetS2}
\end{align}
with 
\begin{align}
    S_{m,m^\prime}(k, k-\Delta k) & := \langle \phi_{m, k} \vert \exp{(\mathrm{i} \frac{2\pi}{L}x)}\vert
    \phi_{m^\prime, k - \Delta k}\rangle \notag \\ 
    & = \langle u_{m, k} \vert u_{m^\prime, k-\Delta k} \rangle. 
    \label{eqDetS3}
\end{align}
The $u_{m, k}$ is the cell-periodic part of the Bloch function. The discrete Wilson line is precisely defined as the overlap matrix between neighboring discrete points in $k$-space, and the discrete Wilson loop can be expressed by 
\begin{align}
    W = \prod_k \mathcal{M}^{(k)}, \quad \text{with } \mathcal{M}^{(k)}_{m,m^\prime} = \langle u_{m, k} \vert u_{m^\prime, k-\Delta k} \rangle.
    \label{eqDetS4}
\end{align}
Comparing the equations (\ref{eqDetS}),(\ref{eqDetS2}),(\ref{eqDetS3}) and (\ref{eqDetS4}), we can see that Resta's definition (\ref{eqRestaDef}) is essentially another expression of the electronic polarization formula $e\langle x \rangle /L :=P_\mathrm{el}=\frac{e}{2\pi}\mathrm{Im}\ln \det W$ in mordern polarization theory.

In his original discussion about the position operator, Resta claimed that the quantity related to the Berry phase (polarization) cannot be written as the expectation value of any operator. This conclusion sparked some debate: Zak argued that this conclusion does not necessarily hold for the geometric phase of energy bands and provided an explicit Hermitian position operator that allows the result to indeed be written as an expectation value \cite{zak2000}. In fact, the lack of a simple, well-defined position operator is inconvenient, as it makes the discussion obscure, especially in the study of localization problems \cite{resta1999}. 
Therefore, when focusing on the study of localization, a simpler computational scheme is needed. 

A simple re-definition of the position operator in PBC is given recently by Berger and co-workers \cite{valencaferreiradearagao2019}:
\begin{equation}
\hat{q}_L(x) :=\frac{L}{2\pi\mathrm{i}}\left[\exp(\frac{2\pi\mathrm{i}}{L} \hat{x}) - 1 \right].
\label{eqBergerDef}
\end{equation}
This redefined operator has the following important properties: 1. periodicity; 2. correct reduction in the appropriate limits; 3. gauge invariance of the distance;
4. being single-particle operator.
Therefore, it is self-consistent and very convenient for calculating higher-order cumulants. It is quite natural to obtain the redefined $\hat{X}$ operator through the complex position (\ref{eqBergerDef}) 
\begin{equation}
\hat{X} = \sum_{i=1}^N x_i \to \hat{Q} = \sum_{i=1}^N \hat{q}_L(x_i).
\label{eqQ}
\end{equation}

There were different ways to  redefine the position operator \cite{zak2000}, but subsequently Berger and co-workers demonstrated the redefined position operator (\ref{eqBergerDef}) is unique under certain evident constraints in quantum physics \cite{evangelisti2022}. From a deeper perspective, the definition in equation (\ref{eqBergerDef}) is a kind of cumulant \cite{hetenyi2025, hetenyi2024}, which can be obtained by treating equation (\ref{eqRestaDef}) as a generating function \cite{hetenyi2019, keralavarma2015}. The redefining (\ref{eqBergerDef}) has been applied recently in the calculations of ionic crystals \cite{tavernier2020,tavernier2021}, Wigner crystals \cite{azor2021,alves2021,escobarazor2019, escobarazor2024} and other quantum chemistry systems with non-trivial geometry \cite{brumas2024, alrakik2025}. To date, although there are various methods to resolve the problem of the position operator in PBC \cite{ditler2023}, the scheme of redefining the position operator is the simplest and widely used \cite{angeli2021,evangelisti2022}. 

The most important application of the redefining position operator is the generalization of the \textit{localization tensor}, which is essentially the spread of the redefined position operator. This generalization facilitates the discussion of localization and dimerization properties in systems with various of boundary conditions \cite{angeli2021,azor2021,tao2022,gong2023}. Resta’s original discussion of the polarization problem was based on a semiclassical physical picture \cite{resta1994}, however, localization is actually correlation information \cite{resta1999}, as the density distribution in multi-electron systems tends to be uniform. The localization tensor precisely provides the correlation information we need, making it a crucial indicator of localization in a system.

Although the position operator defined by equation (\ref{eqBergerDef}) has demonstrated certain advantages in the study of localization, it also has obvious drawbacks. Firstly, this redefined operator loses its Hermiticity, which means that it is not a well defined physical observable \cite{griffiths2018,landau2013}. Additionally, equation (2) only converges to the conventional definition of the position operator as $L \to \infty$, and in general cases, it does not clearly convey `positional' information. More critically, the localization tensor sometimes cannot provide an accurate criterion for localization. For example, in the work where the concept of the localization tensor was formally introduced \cite{valencaferreiradearagao2019}, the authors were actually studying the physical phenomenon of dimerization.

To address the problems, in this paper, we will investigate the problem of localization using a new approach. Our main innovations include the following:
(i) providing a new interpretation of localization from the perspectives of geometry and probability theory;
(ii) proposing a generalization under periodic boundary conditions (PBC) using the Riemannian center and mutual information, which does not require a redefinition of the position operator;
(iii) revealing that the localization tensor (LT) is determined by the asymptotic behavior of the static structure factor near $k=0$, and presenting a criterion and upper bound for distinguishing localization from dimerization---an aspect overlooked in previous studies \cite{valencaferreiradearagao2019, tao2022, gong2023}.

\section{\label{secRedefineAndLT} The localization tensor}


Based on the previous series of studies on the localization of electrons \cite{kohn1964,resta1998,resta1999},  the localization tensor has been proposed as the total position spread per electron \cite{valencaferreiradearagao2019}, explicitly
\begin{equation}
\lambda(N) = \frac{1}{N}[\langle \hat{X}^2 \rangle - \langle \hat{X} \rangle^2].
\label{eqLT}
\end{equation}
Note that (\ref{eqLT}) is defined in the open boundary conditions (OBCs), or in other words, on the flat manifold $\mathbb{R}^n$. By redefining position operator, it can be extended to the case of PBCs as
\begin{equation}
\lambda_L(N) = \frac{1}{N}\left[ \langle \Psi \vert \hat{Q}^\dagger \hat{Q} \vert \Psi \rangle  - \vert\langle\Psi \vert \hat{Q} \vert \Psi  \rangle\vert^2 \right],
\label{eqLTP}
\end{equation}
here the $\hat{Q}$ is the redefined position operators, see (\ref{eqQ}). This definition can also be regarded as a second-order cumulant, reflecting the correlation information of the systems \cite{hetenyi2019, hetenyi2024, hetenyi2025, keralavarma2015}.
Recent studies have indicated that when investigating localization problems, the localization tensor (LT) defined in (\ref{eqLTP}) is a better index compared to traditional ones \cite{gong2023}, being more sensitive to Anderson transitions \cite{tao2022}.

\section{\label{secLocalizationTensor}Localization tensor for general manifold}

In this section, we will present the main results of this work. As we introduced in Sec.\ref{sec:introduction}, the extension from LT in OBCs (\ref{eqLT}) to LT in PBCs (\ref{eqLTP}) has an apparent artificial aspect, as it relies on the redefinition of the position operator (\ref{eqQ}), but this redefinition is not unique and lacks sufficient physical justification.

Our approach is completely different from previous ones. We first note that in OBC the LT can be expressed as the covariance of two random variables, the joint distribution of these two random variables is entirely determined by the density-density correlation function. Thus, the analysis of the localization tensor can be fully transformed into the covariance of two random variables. The definition of the correlation function is clear for any geometric background, and to compute the covariance, one only needs to specify how the distance is defined. Therefore, in this way, the localization tensor can be directly generalized to any geometric shape and boundary condition.

To correspond with previous works, we first consider the one-dimensional lattice case
\begin{equation}
    \hat{X} = \sum_{m \in \Omega} m\hat{c}^\dagger_m\hat{c}_m ,
    \label{eqXDisDef}
\end{equation}
here $\Omega = \{1, 2, \cdots, N\}$ denoting the sites in the system, $\hat{c}^\dagger_m (\hat{c}_m)$ is the creation (annihilation) operator on the $m$-th site. 
We further assume the system conserves the number of particles $\sum_{m \in \Omega} \,\hat{c}^\dagger_m\hat{c}_m = M $ with occupied number $M$ (systems that do not conserve particle number can be extended without much difficulty). By the equation (\ref{eqXDisDef}) and the definition of LT in (\ref{eqLT}), we write the LT in the form of the covariance
\begin{align}
    \lambda(N) &= \frac{M^2}{N} \left(\mathrm{E}[X_1X_2] -\mathrm{E}[X_1]\mathrm{E}[X_2] \right) \notag \\
    &= \frac{M^2}{N}\mathrm{Cov}[X_1, X_2],
    \label{eqCovariance}
\end{align}
here $X_1$ and $X_2$ are random variables, $\mathrm{E}[X_{1,2}]$ denotes the mean value of $X_{1,2}$ and $\mathrm{Cov}[X_1,X_2]$ denotes the covariance between $X_1$ and $X_2$. The distribution function of $X_1$ and $X_2$ is defined via the density-density correlation function
\begin{equation}
    f(x_1,x_2) := \frac{1}{M^2}\langle  \hat{c}^\dagger_{x_1} \hat{c}_{x_1}\hat{c}^\dagger_{x_2} \hat{c}_{x_2}  \rangle = \frac{1}{M^2}\langle \hat{n}_{x_1} \hat{n}_{x_2} \rangle. 
    \label{eqDistribution}
\end{equation}
The positive definiteness of the density-density correlation together with the constrain $\sum_{m \in \Omega} \hat{c}^\dagger_m \hat{c}_m = M$ ensures that $f(x_1,x_2)$ in (\ref{eqDistribution}) is well-defined as a density distribution.

In quantum many-body systems, the density-density correlation function has always played an important role. For a long time, it has been used to describe the correlations between particles \cite{fetter2012}, investigate phase transitions and critical phenomena \cite{baus1983,tsvelik2003}, calculate physical quantities such as the structure factor, compressibility and dielectric constant \cite{chaikin2000}, and reveal long-range order and disorder \cite{wen2007}. The peculiarity of equations (\ref{eqCovariance}) and (\ref{eqDistribution}) lies in the fact that, unlike previous studies that characterized the system through the behavior of density-density correlations between two points \cite{fetter2012,baus1983,tsvelik2003,chaikin2000,wen2007}, the localization tensor depends on the overall information of the density-density correlation. This probabilistic interpretation makes explicit that LT fundamentally measures how strongly two electrons’ positions fluctuate together, rather than the spread of any single‑particle orbital. It is more akin to a statistical characteristic.

The equation (\ref{eqCovariance}) and (\ref{eqDistribution}) were derived in the $\mathbb{R}^n$ case (OBC). Based on the aforementioned understanding of the localization tensor, we will consider extending the localization tensor to the case of PBC. We will extend the LT directly through a geometric perspective in section \ref{secRiemann}, or by interpreting the LT as a statistical quantity from a probabilistic perspective in section \ref{secCrossEntropy}. More importantly, we will reveal the intrinsic relationship between the localization tensor and the correlation function in section \ref{secEssence}, introducing the concept of the localization function. This will help us better understand localized systems.

\subsection{\label{secRiemann}Geometric method: Riemannian centroid}
The definition of equation (\ref{eqCovariance}) is written in the form of moments of random variables. In Euclidean space, the form of moments is explicit \cite{feller1957}, but on a general manifold (for example, the case of $\mathbb{S}^1$ that we are particularly concerned with), there is no explicit definition of moments. This is because the `distance' is a trivial concept in Euclidean space, whereas on a general manifold, it needs to be defined through the additional structure of the Riemann metric tensor and geodesics. 

Fortunately, when we write down the Schr\"{o}dinger equation on an arbitrary manifold, we have implicitly included the structure of the metric tensor \cite{morchio2007,lychagin1999}. For instance, in the $\mathbb{S}^1$ manifold (PBC in one dimension) when parameterized by $\theta \in [0,2\pi)$, the Riemann metric tensor is given by $g = \mathrm{d}\theta^2 = \mathrm{d}\theta \otimes \mathrm{d}\theta$, and the distance between two points $\theta_1$ and $\theta_2$ is given by the minimal length of geodesic curves connecting them:
\begin{equation}
    d(\theta_1,\theta_2) = \begin{cases} 
 2\pi - \vert \theta_1 - \theta_2 \vert& \text{if } \vert \theta_1 - \theta_2 \vert > \pi ; \\
\vert \theta_1 - \theta_2 \vert & \text{if } \vert \theta_1  - \theta_2 \vert  \leq \pi .
\end{cases}
\label{eqDistance}
\end{equation}
Given the defined distance (\ref{eqDistance}), it is straightforward to define the mean value and variance of a random variable on $\mathbb{S}^1$ through the concept of Riemannian center, which has been widely studied and applied in computational geometry \cite{karcher1977,mancinelli2023}. Suppose $\Theta$ is a random variable on $\mathbb{S}^1$ with distribution function $f(\theta)$, then consider the  optimization problem
\begin{align}
    \mathrm{Var}[\Theta] := \min_{\theta_c \in \mathbb{S}^1} \int \mathrm{d}\theta\, f(\theta) \vert d(\theta, \theta_c) \vert^2.
    \label{eqRC}
\end{align}
The minimized value of R.H.S. of (\ref{eqRC}) gives the variance of $\Theta$, while the solution $\theta_c$ who minimized the variance gives the mean value $\mathrm{E}[\Theta]$. This mean value is usually referred as Riemannian center. Although (\ref{eqRC}) is defined in a continuous probability space, its discrete version is easily obtained: simply replace $\pi$ with $N/2$ in the distance definition (\ref{eqDistance}), and change the integral in (\ref{eqRC}) to a summation. Unlike the operator‑redefinition approach, this construction depends only on the underlying metric of the manifold and the density–density correlations, and thus remains valid for any geometry where a geodesic distance can be defined.

Notice that the equation (\ref{eqCovariance}) can be expressed in the form of variance
\begin{align}
    \lambda(N) &= \frac{M^2}{N} \left(\mathrm{E}[X_1X_2] -\mathrm{E}[X_1]\mathrm{E}[X_2] \right) \notag \\  = &\frac{M^2}{2N} \left( \mathrm{Var} [X_1] + \mathrm{Var} [X_2] - \mathrm{Var} [X_1-X_2]\right),
    \label{eqLTRiemann}
\end{align}
and with the help of (\ref{eqRC}), we have obtained specific method for calculating the localization tensor on $\mathbb{S}^1$. As long as we can obtain the probability density distribution defined by (\ref{eqDistribution}), calculating the localization tensor using (\ref{eqLTRiemann}) is straightforward. Clearly, this extension to the PBC case is mathematically rigorous. In the examples in the next section, we will demonstrate the effectiveness of this extension.

\subsection{\label{secCrossEntropy}Statistical method: mutual information}
We can also examine equation (\ref{eqCovariance}) from statistical perspective. We know that the significance of covariance is to measure the correlation between two random variables, so the localization tensor can actually be seen as a criterion of the correlation of the bivariate random distribution (\ref{eqDistribution}). This criterion of correlation is more justifiably standard in probability theory, namely by calculating mutual information. Compared to covariance, the mutual information, as a purely probabilistic quantity \cite{cover2006}, only requires the manifold to be the sample space of the probability space without needing additional structures (such as distance) on the manifold, and therefore can be directly extended to general manifolds. The mutal information is defined by
\begin{equation}
    I[\vert \psi \rangle] = \sum_{x_1, x_2 \in \Omega} f(x_1, x_2) \log\left( \frac{f(x_1,x_2)}{f(x_1)f(x_2)}\right),
    \label{eqMI}
\end{equation}
here for a given wave function $\vert \psi \rangle$, the density distribution $f(x_1,x_2)$ is determined by (\ref{eqDistribution}).

Mutual information, as an indicator, is not as sensitive to localization phase transitions as the localization tensor. However, it exhibits different behavior in localization and dimerization, revealing information that the localization tensor cannot capture. The meaning of mutual information quantifies the correlation between two random variables within a probability distribution. For a purely localized state, observing an electron on one lattice site has little influence on the probability of observing an electron on another site. Therefore, compared to free case, the mutual information will be smaller. In contrast, for a dimerized state, observing an electron on one site tells  that the probability of finding an electron on odd or even sites is different. This odd/even distinction actually constitutes additional information, which increases the mutual information. These characteristics will be demonstrated in the examples in the next section.

\subsection{\label{secEssence}Essence of LT: the asymptotic behavior of the structure factor}
\subsubsection{The SSH model: dimerization}
To clarify the discussion, we return to the model where the localization tensor was originally introduced by E.~V.~F.~de Arag\~ao \emph{et al.} \cite{valencaferreiradearagao2019}. In this work,  the localization tensor was clearly introduced in the study of the Su-Schrieffer-Heeger(SSH) model \cite{su1979,lieu2018} . Although this model is not a typical localization model in the conventional sense, the localization tensor provides a clear criterion for dimerization. Moreover, since the model is exactly solvable, it is very helpful for further discussion.

Consider a periodic chain containing $N$ (even) atoms at half filling, the tight-binding Hamiltonian is given by 
\begin{equation}
    \hat{H}_{\mathrm{d}} = \sum_{i=1}^{N-1} - t_i \hat{a}_i^\dagger \hat{a}_{i+1} - t_N \hat{a}_N^\dagger \hat{a}_1 + \mathrm{h.c.},
    \label{eqHd}
\end{equation}
where $\hat{a}_i^\dagger$($\hat{a}_i$) is a creating (annihilation) operator and the hopping parameter $t_i = 1 - \delta (-1)^i$ with $0 \leq\delta\leq 1$. The $\delta$ is the dimerization parameter. The two-point correlation functions for ground state can be given by
\begin{equation}
    \langle \hat{a}_m^\dagger \hat{a}_n \rangle = \frac{1}{N} \sum_{p\in\Omega_f}  \mathrm{e}^{\frac{2\pi\mathrm{i}}{N}(n-m)p}[c_p - \mathrm{i}(-1)^m s_p][(c_p + \mathrm{i}(-1)^n s_p],
    \label{eqCor2}
\end{equation}
here $\Omega_f = \{0, 1, \cdots, N/2-1\}$ denotes the occupied indices in the Fermi sea, $c_p$ and $s_p$ is defined by $c_p := \cos \frac{\theta_p}{2}$ and $s_p := \sin \frac{\theta_p}{2}$ with $\tan\theta_p := \lambda \tan( \frac{2\pi}{N}p)$. Because (\ref{eqHd}) describes free fermionic system, the density-density correlation can be evaluated by Wick's theorem
\begin{align}
    \langle \hat{n}_m \hat{n}_n \rangle = \langle \hat{a}_m^\dagger \hat{a}_m\rangle \langle \hat{a}_n^\dagger \hat{a}_n\rangle+ \langle \hat{a}_m^\dagger \hat{a}_n \rangle \langle \hat{a}_m \hat{a}_n^\dagger \rangle.
\end{align}
Based on the density-density correlation function, we can define the \textit{static structure factor} 
\begin{equation}
    \tilde{C}_{pq} = \frac{1}{N}\sum_{m,n} \mathrm{e}^{\frac{2\pi\mathrm{i}}{N}(nq-mp)}C_{mn}
    \label{eqCpq}
\end{equation}
with correlator $C_{mn} := \langle \hat{a}_m^\dagger \hat{a}_m \hat{a}_n^\dagger \hat{a}_n \rangle - \langle \hat{a}_m^\dagger \hat{a}_m\rangle \langle \hat{a}_n^\dagger \hat{a}_n\rangle$. The static structure factor is a fundamental quantity in condensed matter physics, with a broad experimental foundation. It serves as a central focus in experiments such as X-ray scattering\cite{greenfield1971}, neutron scattering\cite{alexandru2021, honecker2020}, and electron diffraction\cite{filippetto2022, gemmi2019}. Remark that in conventional studies of condensed matter physics, translational invariance of the system ensures the translational invariance of the correlator, typically resulting in the static structure factor having only one wave vector degree of freedom \cite{tsvelik2003, wen2007}. However, in our current research, particularly localized systems, this invariance is absent, so the static structure factor has two independent wave vectors. 

Substitute the redefined position operator (\ref{eqBergerDef}) into the definition of the localization tensor (\ref{eqLTP}), it is not hard to show that the localization tensor can be expressed in the form of static structure factor
\begin{equation}
    \lambda(N) = \frac{N^2}{4\pi^2}\left(\tilde{C}_{11} - \tilde{C}_{01} - \tilde{C}_{10} + \tilde{C}_{00}\right).
    \label{eqLTinC}
\end{equation}
Furthermore, according to the exact solution of model (\ref{eqHd}), we can prove $\tilde{C}_{01} = \tilde{C}_{10} = \tilde{C}_{00} = 0$, thus the localization tensor is equal to the  first non-zero diagonal element of the structure factor $\tilde{C}_{pq}$. 

We denote the diagonal elements of 
$\tilde{C}_{pq}$ as 
\begin{equation}
    \tilde{C}_p := \tilde{C}_{pp},
    \label{eqDefLTF}
\end{equation}
in the thermodynamic limit $N \to \infty$, we can obtain analytical expression
\begin{equation}
    \tilde{C}(p)=\frac{1}{4}-\int_0^{2\pi}\frac{\mathrm{d}p'}{8\pi}
    \frac{\cos p + a\cos p'}{\sqrt{(\cos p+ a\cos p')^2+(1-a^2)\sin^2p}},
    \label{eqLF}
\end{equation}
with $a = (1-\delta^2)/(1+\delta^2)$. We name $\tilde{C}(p)$ the \textit{localization function}, and it is evident that the localization tensor in the thermodynamic limit is determined by the localization function via 
\begin{equation}
    \lim_{N \to \infty} \lambda(N) = \frac{1}{2}\frac{\mathrm{d}^2 \tilde{C}(p)}{\mathrm{d}p^2} \Big{\vert}_{p=0}.
    \label{eqLT2d}
\end{equation}
The SSH model (\ref{eqHd}) is critical at $\delta = 0$ where the gap is closed. According to (\ref{eqLF}) it is easy to conclude $\tilde{C}(p)\vert_{\delta = 0} = (2\pi)^{-1}\vert p \vert$, then  (\ref{eqLT2d}) indicates the divergent behavior of the localization tensor in the thermodynamic limit at $\delta = 0$, consistent with the conclusions in \cite{valencaferreiradearagao2019}. To conclude, the localization tensor is determined by the asymptotic behavior of the localization function, which is the diagonal part of the static structure factor, at $p \to 0$. For SSH model,
\begin{equation}
    \tilde{C}(p \to 0^+) = 
    \begin{cases}
        (2\pi)^{-1}p, & \delta=0, \text{ free fermion}; \\
        \lambda(\infty)p^2, & \delta \neq 0, \text{ dimerization}.
    \end{cases}
    \label{eqLFSSH}
\end{equation}

Equation (\ref{eqLFSSH}) indicates that $\tilde{C}(p)$ exhibits distinctly different asymptotic behaviors in different phase regions in SSH model. This phenomenon is also observed in models with localization phase transitions \cite{gong2023,tao2022}, underlying the reason why the localization tensor can serve as an indicator of localization phase transitions. 

\subsubsection{The localization function in the localized phase}
It is important to note the detail that in studies of localization, the localization tensor used is renormalized by factor $N^{-1}$. The localization tensor as defined in this way is actually proportional to the coefficient of the linear term of $\tilde{C}(p \to 0)$ (rather than the coefficient of the quadratic term as in (\ref{eqLFSSH})). This renormalization is intended to ensure that the localization tensor equals $1$ in the extended phase and $0$ in the localized phase.

Next, we present a rough but insightful estimate to analyze the behavior of the localization function within a localized phase. In the localized phase, we assume that the many-body wave function is composed of $M$ localized one-body wave functions $\psi_k(m - x_k)$ by filling the Fermi sea, here $k \in \Omega_f = \{1,2,\cdots,M \}$ is the index in the Fermi sea and $x_k$ denotes the center of the wave package of the $k$-th wave function. Denoting $\rho_k(m) = \psi_k^*(m)\psi_k(m)$ the density distribution of the $k$-th wave function, and further assuming $1 \ll M \ll N$, now we can safely ignore the overlap between different filled one-body wave functions and estimate the correlator $C_{mn} := \langle \hat{a}_m^\dagger \hat{a}_m \hat{a}_n^\dagger \hat{a}_n \rangle - \langle \hat{a}_m^\dagger \hat{a}_m\rangle \langle \hat{a}_n^\dagger \hat{a}_n\rangle$ as
\begin{equation}
    C_{mn} \approx \sum_{k \in \Omega_f}  \delta_{mn}\rho_k(m - x_k) - \rho_k(m - x_k)\rho_k(n - x_k).
    \label{eqCmnApp}
\end{equation}
Define $\tilde{\rho}_k(p) := \sum_m \mathrm{e}^{\frac{2\pi\mathrm{i}}{N}mp} \rho_k(m)$ to be the Fourier transformation of $\rho_k$, then we can give an estimation of the localization function
\begin{equation}
    \tilde{C}(p) \approx \frac{1}{N} \sum_{k \in \Omega_f} \left( 1 - \vert \tilde{\rho}_k(p)\vert^2\right).
    \label{eqCpApp}
\end{equation}
Based on the definition of $\tilde{\rho}_k$ and the normalization condition of the wave function, it is evident that $\tilde{\rho}_k(0) = 1$ for all $k$, thus (\ref{eqCpApp}) ensures that $\tilde{C}(0) = 0$. Furthermore, assuming that $\rho_k(m)$ is a narrow wave package---a natural assumption in the localized phase---its Fourier transform will be a broad wave packet, which in the limit case degenerates to a constant equal to $1$. A typical behavior of $\tilde{\rho}_k(p)$ is that it is a periodic function which monotonically decreases as $p$ increases, until it reaches its minimum at $p \to N/2$ (or $p \to \pi$ in the continuous case). Thus, we can obtain a conservative estimate regarding $\vert \tilde{\rho}_k(p) \vert \geq \vert \tilde{\rho}_k(\pi) \vert \geq \sqrt{2R_k - 1}$, here the $R_k$ is the inverse partition ratio (IPR) of the $k$-th wave function in the Fermi sea, defined by
\begin{equation}
    R_k := \sum_m \vert \psi_k(m) \vert^4 / \sum_m \vert \psi_k(m) \vert^2.
    \label{eqIPR}
\end{equation}
The IPR is a crucial metric for studying the characteristics of wave function localization, with a tendency towards $1$ corresponding to complete localization. According to the preceding arguments, we can derive an upper bound for the localization function provided by the IPR
\begin{equation}
    \tilde{C}(p) \leq \frac{1}{N}\sum_{k \in \Omega_f} (2 - 2R_k).
    \label{eqCpBound}
\end{equation}
The bound (27) shows that in a deeply localized phase, where each occupied orbital has $\mathrm{IPR} \to 1$ , the localization function  is suppressed at all finite momenta, not only near $p =0$.
Numerical calculations indicate that the bound in (\ref{eqCpBound}) is relatively tight when $M$ is much smaller than $N$, and although it remains valid as $M$ increases, it becomes looser. Obviously we have $\tilde{C}(p) \to 0$ when $R_k \to 1$, corresponding to the deep localized case. 

We summarize the behavior of the localization function $\tilde{C}(p)$ in the localized phase: when $p \to 0^+$, according to (\ref{eqCpApp}), the $\tilde{C}(p)$ behaves as a quadratic function, contrasts sharply with its linear behavior in the extended phase. Analogous to (\ref{eqLFSSH}), we have
\begin{equation}
    \tilde{C}(p \to 0^+) \propto
    \begin{cases}
        p, &\text{extended phase}; \\
        p^2, & \text{localized phase}.
    \end{cases}
    \label{eqLFAA}
\end{equation}
The result (\ref{eqLFAA}) explains why the localization tensor is a highly sensitive indicator for localization phase transitions. 

\subsubsection{Distinguish localization and dimerization}
An important but previously overlooked fact is that the localization tensor is determined only by the asymptotic behavior of the localization function near zero. Therefore, according to Equations (\ref{eqLFSSH}) and (\ref{eqLFAA}), the localization tensor cannot distinguish between localization and dimerization. Hence, as an indicator of localization, it is fully incomplete. To accurately identify localization or dimerization, we require more comprehensive information about the localization function. Notably, in the localized phase, the localization function is subject to an upper bound given by (\ref{eqCpBound}). In contrast, for the case of dimerization, equation (\ref{eqLF}) dicates that $\tilde{C}(p)$ strictly equals $1/2$ when $p = \pi$. Therefore, a crucial distinction between localization and dimerization lies in the behavior as $p \neq 0$, typically when $p \to \pi$.

\begin{figure}
\centering
\begin{center}
\includegraphics[width=0.98\linewidth]{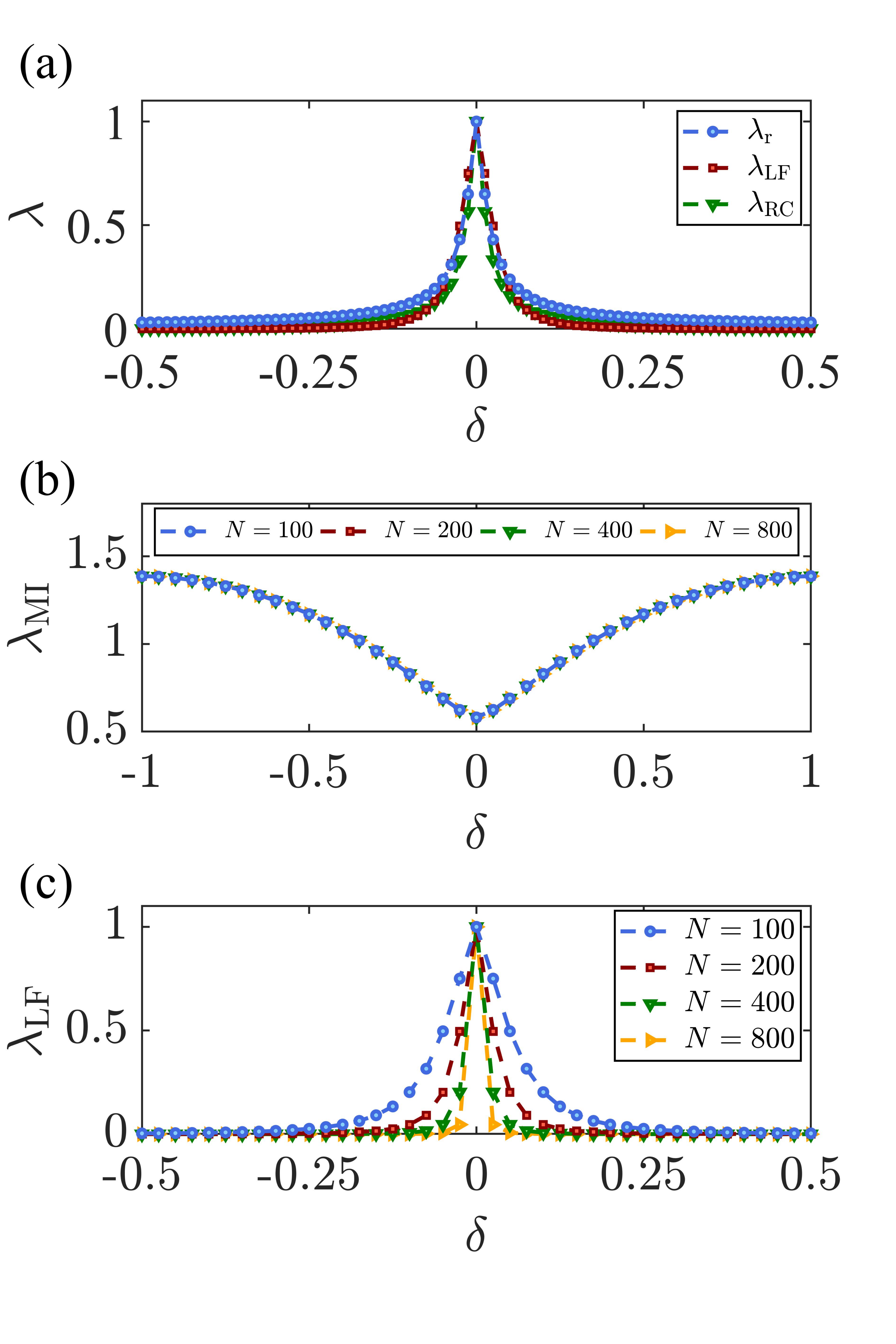}
\end{center}
\caption
{The various definitions of the localization tensor $\lambda$ for different values of the dimerization parameter $\delta$ in the Su-Schrieffer-Heeger model (\ref{eqHd}) are shown. (a) We set $N = 200$ and $M = N/2$ (half filling), using different line styles to distinguish between the definitions $\lambda_r$, $\lambda_{\mathrm{LF}}$, and $\lambda_{\mathrm{RC}}$. (b) Different line styles represent the results for $\lambda_{\mathrm{MI}}$ with varying $N$. (c) The results for $\lambda_{\mathrm{LF}}$ with different $N$.}
\label{fig-LT_lambda}
\end{figure}

\section{\label{secExample}Examples}
In the previous section, we conducted a detailed analysis of the physical essence of the localization tensor and proposed various methods for its generalization, here we summarize these approaches. We have proposed the following indicators as extensions of the localization tensor:
\begin{align}
    \lambda_\mathrm{r} &=  \frac{N^2}{4\pi^2}\left(\langle \Psi \vert \hat{Q}^\dagger \hat{Q} \vert \Psi \rangle  - \vert\langle\Psi \vert \hat{Q} \vert \Psi  \rangle\vert^2 \right) \notag \\
    \lambda_\mathrm{RC} &= \frac{N^2}{4\pi^2} \left( \mathrm{Var} [X_1] + \mathrm{Var} [X_2] - \mathrm{Var} [X_1-X_2]\right)  \notag \\
    \lambda_\mathrm{MI} &= N\sum_{x_1, x_2 \in \Omega} f(x_1, x_2) \log\left( \frac{f(x_1,x_2)}{f(x_1)f(x_2)}\right) \notag \\
    \lambda_\mathrm{LF} &= 2\pi \frac{\mathrm{d} \tilde{C}(p)}{\mathrm{d}p} \Big{\vert}_{p=0}. \notag
\end{align}
The definition of $\lambda_\mathrm{r}$ is based on the original definition of $\lambda_L(N)$, as seen in (\ref{eqLTP}), notice that we have added a normalization factor to ensure its value is exactly $1$ in the case of free fermions. The $\lambda_\mathrm{RC}$ and $\lambda_\mathrm{MI}$ are defined referring to  (\ref{eqLTRiemann}) and (\ref{eqMI}), with distribution function $f(x_1,x_2)$ obtained by (\ref{eqDistribution}) and $X_1,X_2$ are random variables determined by $f(x_1,x_2)$. The $\lambda_\mathrm{LF}$ is defined through the asymptotic behavior of the localization function $\tilde{C}(p)$, which is obtained through the continuous extension of the diagonal part in (\ref{eqCpq}). Similarly, we have added the normalization factor to ensure that its value is exactly $1$ in the case of free fermions.

\begin{figure}[t]
\centering
\begin{center}
\includegraphics[width=0.98\linewidth]{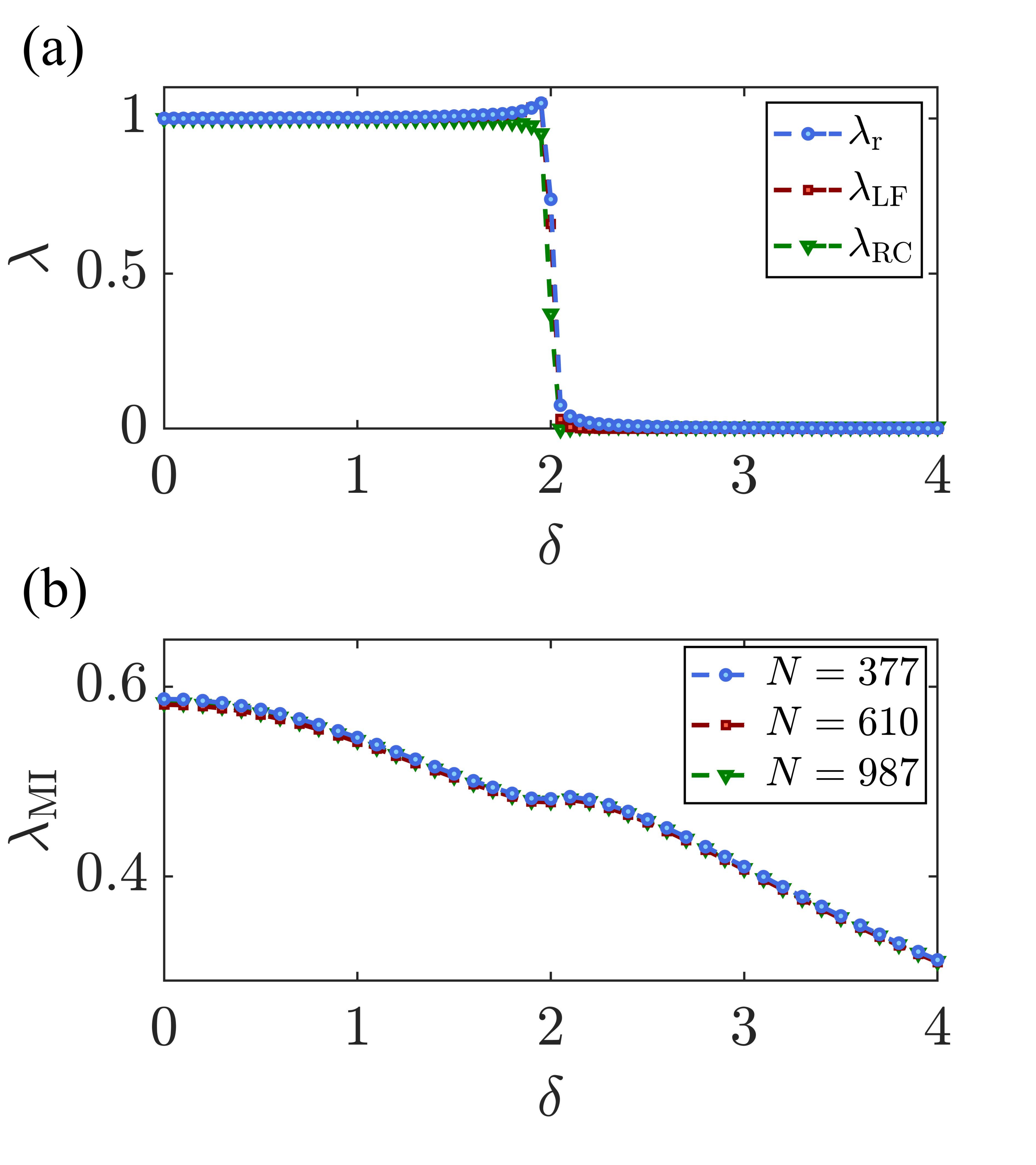}
\end{center}
\caption
{The different definitions of localization tensor $\lambda$ for various values of the localization parameter $\delta$ in Aubry-Andr\"e model (\ref{eqHAA}). We take $\beta=(\sqrt{5}-1)/2$, $V=0$ (non-interacting) and $J=1$. (a) Here $N=610$, filling $M=N/2$, different types of lines are used to distinguish different definitions of $\lambda_{r}$, $\lambda_{\rm LF}$ and $\lambda_{\rm RC}$. (b) The results of $\lambda_{\rm MI}$ under different $N$.}
\label{fig-AA}
\end{figure}

\subsection{\label{LT_D-L}Localization tensor for extended–dimerized and extended–localized phase transitions}
To investigate the characteristics of these metrics above, we focus on two specific models. One is the model previously discussed with Hamiltonian (\ref{eqHd}), with a particular emphasis on dimerization behavior; the other is the Aubry-Andr\"e (AA) model 
\begin{align}
H = &\sum_{i=1}^{N}J(\hat{a}_i^\dagger\hat{a}_{i+1}+\mathrm{h.c.}) + \delta\cos(2\pi\beta i)\hat{a}_i^\dagger\hat{a}_i \notag \\
&+V\hat{a}_i^\dagger\hat{a}_i\hat{a}_{i+1}^\dagger\hat{a}_{i+1}
\label{eqHAA}
\end{align}
which is extensively used in studies of localization\cite{biddle2009,iyer2013,huse2014,tao2022}. Here, $J$ is the hopping strength, $\delta$ is the potential strength, and $V$ is the interaction strength between adjacent particles. We have introduced the interacting terms in (\ref{eqHAA}) so that our example can be used to discuss many-body localization as well as Anderson localization.

First, we study the SSH model (\ref{eqHd}), which leads to dimerization phenomena. We use different definitions to calculate the localization tensor $\lambda$ of the system. The results are shown in Fig.\ref{fig-LT_lambda}. We used different types of curves to represent the values of $\lambda_{\rm r}$, $\lambda_{\rm LF}$, and $\lambda_{\rm RC}$ at different $\delta$. In Fig.\ref{fig-LT_lambda}(a), we observe that $\lambda_\mathrm{r}$, $\lambda_{\mathrm{LF}}$, and $\lambda_{\mathrm{RC}}$ are highly sensitive to changes in the dimerization parameter $\delta$ when it is close to zero and reach their maximum value of 1 in the case of free fermions at $\delta = 0$. In Fig.\ref{fig-LT_lambda}(b), we show the dependence of $\lambda_{\mathrm{MI}}$ on $\delta$. This parameter is less sensitive than the other three localization tensors, and its dependence on system size is also less pronounced. Nevertheless, it still reaches its extremum at $\delta = 0$. In the free fermion limit, the density-density correlation, as a bivariate distribution function, exhibits minimal independence between its two variables.
In Fig.\ref{fig-LT_lambda}(c), we show the sensitivity of $\lambda_{\mathrm{LF}}$ to $\delta$ for different system sizes. It can be seen that as the system size increases, the width of the peak in $\lambda_{\mathrm{LF}}$ becomes narrower, indicating a higher sensitivity to dimerization.
 
\begin{figure}
\centering
\begin{center}
\includegraphics[width=0.98\linewidth]{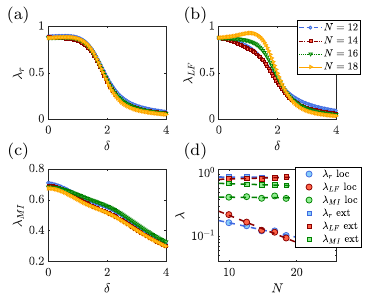}
\end{center}
\caption
{The behavior of several different localization tensors in the ground state of the AA model described by equation (\ref{eqHAA}). Our calculations are based on exact diagonalization and are performed at half filling. We set $J = 1$, $V = 0.25$, $\beta = (\sqrt{5}-1)/2$ and vary $\delta$ to drive the localization phase transition. Panel (a-c) are the three different type of localization tensor discussed in the main text, and Panle (d) shows the scaling behavior to confirm the convergence of the numerical calculations. We choose $\delta = 3.0$ to represent the localized phase while $\delta = 1.0$ to the extended phase.}
\label{fig-MBL}
\end{figure}

In Fig.\ref{fig-AA}, we consider the AA model (\ref{eqHAA}) without interaction,  select an irrational number with phase $\beta=(\sqrt{5}-1)/2$. The system has a quasi disordered periodic potential. Because $V = 0$, the model reduces to the classic case exhibiting the Anderson localization transition. As shown in (a) of Fig. (\ref{fig-AA}),at $\delta = 2$, the system undergoes the Anderson localization transition. We observe that $\lambda_\mathrm{r}$, $\lambda_{\mathrm{LF}}$, and $\lambda_{\mathrm{RC}}$ are highly sensitive to this transition, exhibiting a sharp change at the critical point that resembles a step function. In Fig.\ref{fig-AA}(b), we show the sensitivity of $\lambda_{\mathrm{MI}}$ to $\delta$ for different system sizes. Similarly to the case in the SSH model, this localization tensor is not highly sensitive to the phase transition; it decreases as the value of $\delta$ increases.

In Fig.\ref{fig-MBL}, we consider the interacting AA model, where localization manifests as many-body localization. Due to finite-size limitations (with calculations performed up to $N = 18$), the changes in the localization tensor are much more gradual compared to those in Fig.\ref{fig-AA}. Nevertheless, the transition from the extended phase to the localized phase can still be clearly characterized.

Through the above discussion, we find that the localization tensor can indeed distinguish the transitions between the extended phase and the localized phase, as well as between the extended phase and the dimerized phase. However, when it comes to distinguishing between the localized phase and the dimerized phase, the simple localization tensor does not provide sufficient information. As we will see next, by generalizing the localization tensor to the localization function, the localized and dimerized phases can be clearly identified.

\begin{figure}
\centering
\begin{center}
\includegraphics[width=0.98\linewidth]{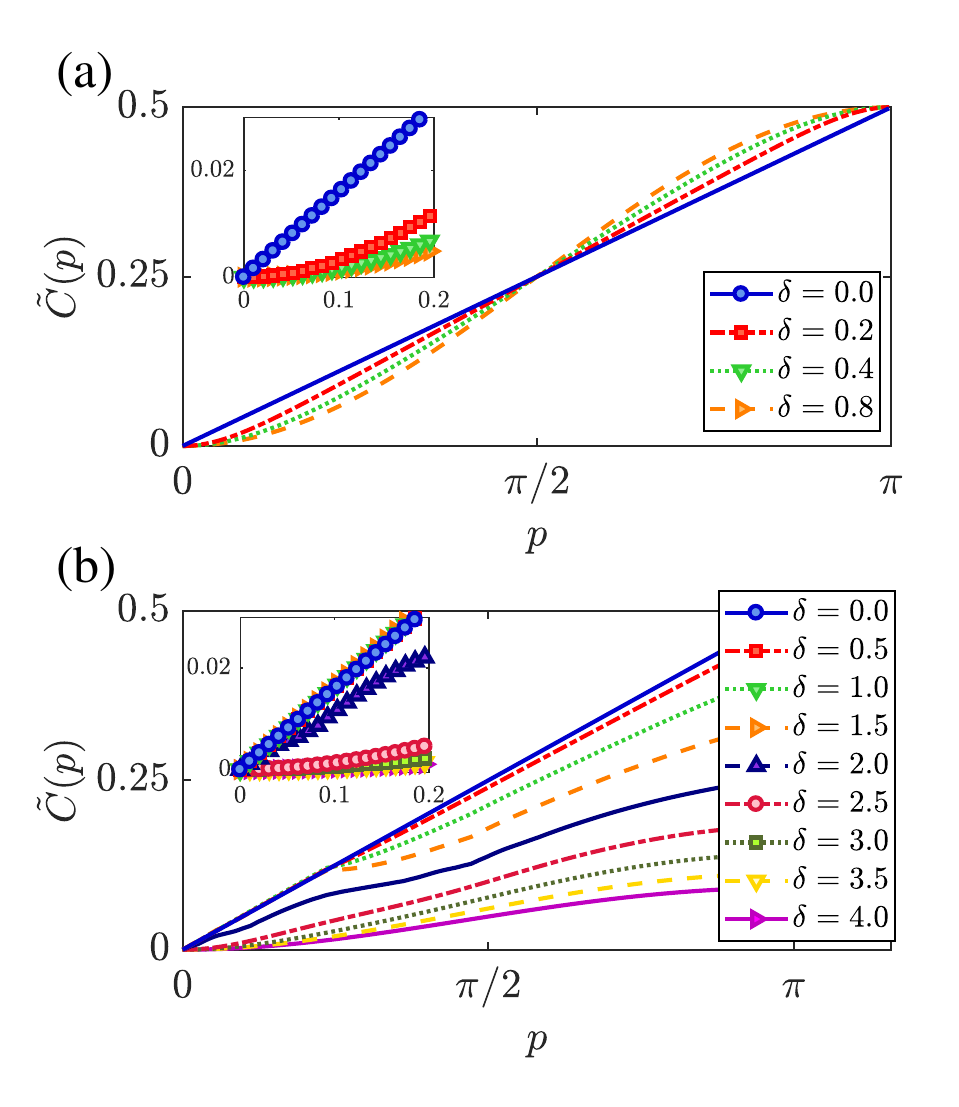}
\end{center}
\caption
{The value of localization function $\tilde{C}(p)$ varying with $\delta$ under different models, and we use different types of lines to distinguish $\tilde{C}(p)$ of different $\delta$. (a) The SSH model (\ref{eqHd}) with the size $N=610$ and $M=N/2$. Inset: The asymptotic behavior of curves approaching $0^+$. (b) The AA model (\ref{eqHAA}) with the size $N=610$ and $M=N/2$. We take $\beta=(\sqrt{5}-1)/2$, $V=0$ and $J=1$. Inset: The asymptotic behavior of $\tilde{C}(p)$ when $p \to 0^+$.}
\label{fig-DL}
\end{figure}

\subsection{\label{LF_D-L}Localization function distinguishing dimerized and localized phases}
In Fig.\ref{fig-DL}, we present the behavior of the localization function, as defined in equation (\ref{eqDefLTF}), in both the SSH model and the AA model as various parameters are varied. The insets show that, when $\delta \neq 0$ in the SSH model and when $\delta > 2$ in the AA model, the behavior of the two models as $p \to 0$ becomes indistinguishable. This is consistent with the conclusions we obtained in (\ref{eqLFSSH}) and (\ref{eqLFAA}). By comparing Fig.\ref{fig-DL}(a) and Fig.\ref{fig-DL}(b), we can clearly see that the behavior of the localization function is completely different in the dimerized phase and the localized phase. In the dimerized case shown in Fig.\ref{fig-DL}(a), the localization function increases monotonically with $p$ and reaches its maximum value of 0.5 at $p = \pi$, indicating a certain even-odd periodicity in the long-range correlations of the system. In contrast, in the localized phase shown in (b), as $p$ increases, it is evident that the value of $\tilde{C}(p)$ is increasingly suppressed as the strength of the disorder potential $\delta$ in (\ref{eqHAA}) increases.

\begin{figure}
\centering
\begin{center}
\includegraphics[width=0.98\linewidth]{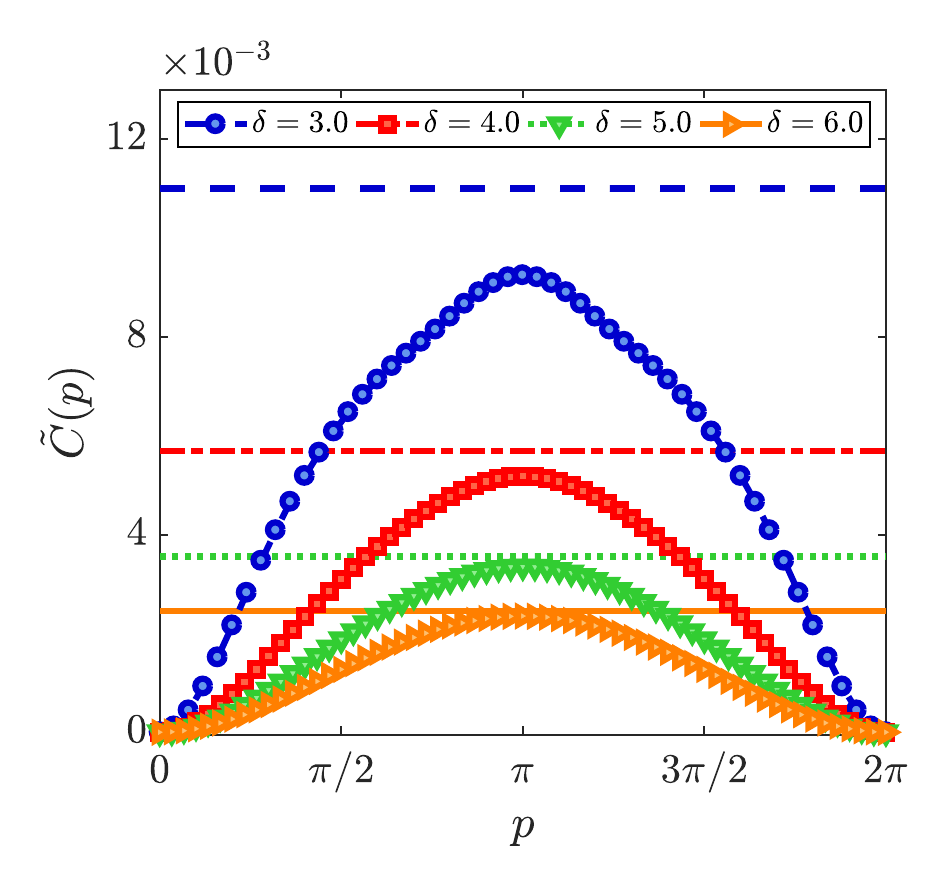}
\end{center}
\caption
{The localization function of AA model (\ref{eqHAA}) under different $\delta$ with the size $N=610$ and $M=20$, and we use different types of lines to distinguish different $\delta$, the curve of $\tilde{C}(p)$ are with graphical markers, while the unmarked lines give the upper bounds(\ref{eqCpBound}) of $\tilde{C}(p)$. Here $\beta=(\sqrt{5}-1)/2$, $V=0$ and $J=1$.}
\label{fig-Cp}
\end{figure}

The suppression of $\tilde{C}(p)$ by the disorder strength can be clearly expressed by the bound we proposed in equation (\ref{eqCpBound}). Note that although $\tilde{C}(p)$ is calculated from the many-body wavefunction, the bound on the right-hand side of equation (\ref{eqCpBound}) depends only on the $\mathrm{IPR}$ values of each energy level. Moreover, this bound decreases monotonically as the degree of localization increases. In Fig.\ref{fig-Cp}, the constraint imposed by the bound on $\tilde{C}(p)$ can be clearly seen, here, the various horizontal lines indicate the bounds corresponding to different disorder strengths. In particular, for smaller filling $M \ll N$, this bound becomes increasingly tight as disorder increases. It is precisely the existence of this bound that leads to the markedly different behaviors of $\tilde{C}(p)$ in the localized and dimerized phases.

In conclusion, In the SSH chain, 
$\tilde{C}(p)$ increases monotonically and reaches a universal plateau 
$\tilde{C}{\pi} \to \frac{1}{2}$, reflecting robust dimer‑scale correlations. In the localized AA chain, 
$\tilde{C}(p)$ is strongly suppressed at finite $p$ as the potential strength grows, consistent with the IPR‑based upper bound. By generalizing the localization tensor to the localization function, we have enabled it to distinguish between localization and dimerization. This also makes its naming more appropriate—otherwise, in many models, including the SSH model where it was originally introduced \cite{valencaferreiradearagao2019}, it does not serve as a genuine indicator of localization.

\section{\label{secSummary}Summary and Outlook}
In this article, we have discussed several strategies to address the challenges associated with the periodicity of position operators under periodic boundary conditions (PBC). From geometric, statistical, and static structure factor perspectives, we have proposed several definitions of the localization tensor. These definitions are motivated by physical considerations and do not rely on a redefinition of the position operator. We have verified their effectiveness in distinguishing both the extended-to-localized and the extended-to-dimerized phase transitions.

Through this framework, we deconstructed the widely used localization tensor and demonstrated that it can be related to the behavior of a localization function, $\tilde{C}(p)$, in the vicinity of $p \to 0^+$ in the thermodynamic limit. The value of the localization function at nonzero $p$ allows us to distinguish between localization and dimerization which renders the localization tensor more logically self-consistent. We have also established an upper bound for this function in the localized regime. As illustrated by the examples presented at the end part of this article, our conclusions appear to be robust and may be broadly applicable. The definitions and results provided here can be readily generalized to a wide range of systems.

Our geometric–probabilistic framework can be extended to non-Hermitian systems in a direct manner and with 
clear practical relevance. Since the construction is based on the spatial statistics of eigenmodes, it does
not require Hermiticity; in the non-Hermitian case one can use the  bi-orthogonal mode weight defined from
left/right eigenstates, and the correlation functions will be well-defined. This enables straightforward 
applications to non-Hermitian models and provides a useful way to separate genuine bulk localization from 
boundary accumulation caused by the non-Hermitian skin effect. The point that requires special 
attention is that, in the non-Hermitian case, the definition of the Brillouin zone may change \cite{alase2017, alase2016, yao2018, yokomizo2019, yang2020}. This means we must be more careful when computing the localization function.

\begin{acknowledgments}
This work is supported by MOST 2022YFA1402701, and the National Natural Science Foundation of China under Grant Nos. 12274419, 12104372 and 12134015, and the CAS Project for Young Scientists in Basic Research under Grant No.YSBR-055.
\end{acknowledgments}





\bibliography{PositionOperator}

\end{document}